\documentclass[aps,showpacs,graphicx,twocolumn]{revtex4}
\usepackage{amsmath}
\usepackage{amscd}
\usepackage{graphicx}
\begin{document}

\title{Efficient polarization entanglement purification based on parametric down-conversion sources with cross-Kerr
nonlinearity\footnote{Published in Phys. Rev. A \textbf{77}, 042308 (2008)}}

\author{Yu-Bo Sheng,$^{1,2}$ Fu-Guo Deng,$^{1,3}$\footnote{Email address:
fgdeng@bnu.edu.cn}  and Hong-Yu Zhou$^{1,2}$}
\address{$^1$The Key Laboratory of Beam Technology and Material
Modification of Ministry of Education, Beijing Normal University,
Beijing 100875, People's Republic of China\\
$^2$Institute of Low Energy Nuclear Physics, and Department of
Material Science and Engineering, Beijing Normal University,
Beijing 100875, People's Republic of China\\
$^3$Department of Physics, Applied Optics Beijing Area Major
Laboratory, Beijing Normal University, Beijing 100875,  People's
Republic of China }
\date{\today }

\begin{abstract}
We present a way for entanglement purification based on two
parametric down-conversion (PDC) sources with cross-Kerr
nonlinearities. It is comprised of two processes. The first one is
a primary entanglement purification protocol for PDC sources with
nondestructive quantum nondemolition (QND) detectors by
transferring the spatial entanglement of photon pairs to their
polarization. In this time, the QND detectors act as the role of
controlled-not (CNot) gates. Also they can distinguish the photon
number of the spatial modes, which provides a good way for the
next process to purify the entanglement of the photon pairs kept
more. In the second process for entanglement purification, new QND
detectors are designed to act as the role of CNot gates. This
protocol has the advantage of high yield and it requires neither
CNot gates based on linear optical elements nor sophisticated
single-photon detectors, which makes it more convenient in
practical applications.
\end{abstract}
 \maketitle

\section{introduction}

Quantum entanglement plays an important role in quantum
information processing and transmission, such as quantum
computation \cite{A. Barenco}, quantum teleportation
\cite{teleportation}, quantum dense coding \cite{densecoding},
quantum state sharing \cite{QSSpra} and certain types of quantum
cryptography \cite{Ekert91,BBM92,rmp,long,two-step,lixhpra}. In
order to complete these tasks efficiently, people need to share
some maximally entangled states. In a practical transmission, the
interaction between a quantum system and the innocent noise of
quantum channel (such as optical fibers or a free space) will
inevitably occur, which will degrade the entanglement of the
quantum system or even make it in a mixed state. The impurity of
the quantum system will make the outcome of the quantum
computation anamorphic, the fidelity of quantum teleportation
degraded, quantum dense coding failed and the key in quantum
cryptography insecure. If the destructive effect of the noise is
not very much, one can exploit entanglement concentration or
entanglement purification to improve the entanglement of the
quantum system first, and then achieve the goals of the
applications above with maximally entangled state.

Entanglement concentration \cite{ec1,ec2,ec3,ec4} is used to
increase the entanglement of some pure entangled pairs at the risk
of that of some others. For the more general case of the quantum
system transmitted through a noisy channel, it is in a mixed state
and the process for reconstructing it in a maximally entangled
state with an ensemble is termed as entanglement purification or
distillation \cite{C. H. Benneet,C. H. Benneet2,D. Deutsch,H.-J.
Briegel,panjw,K. Chen,M. Horodecki,L.-M. Duan,M. Murao,A.
Miyake,Yong}. Generally, the implementation of entanglement
purification schemes requires two or more controlled-not (CNot)
gates which is not experimentally feasible with linear optical
elements at present. In 1996, Bennett \emph{et al.} \cite{C. H.
Benneet} proposed an original entanglement purification scheme for
purifying a Werner state \cite{Werner} with two CNOT gates and
single-photon measurements. Subsequently, Deutsch \emph{et al.}
\cite{D. Deutsch} optimized this scheme for quantum privacy
amplification with two CNOT operations and two special unitary
transforms.

In 2001, Pan \emph{et al.} \cite{panjw} proposed an entanglement
purification protocol with linear optical elements such as
polarizing beam splitters (PBSs) and quarter wave plates (QWPs).
In their protocol \cite{panjw}, the two PBSs are used to complete
the task of parity-check measurements of polarized photons with
their spatial modes. We call it PBS protocol below. This protocol
succeeds, provided that two ideal entangled sources are used. That
is, both  emit one and only one entangled photon pair
synchronously at each time slot. As pointed out by Simon and Pan
\cite{polarization} in 2002,  the currently available source of
entangled photons, parametric down-conversion (PDC), is not an
ideal entangled source.  The feature of PDC seems to fail for the
PBS protocol \cite{panjw}. They then proposed a new entanglement
purification protocol by exploiting spatial entanglement to purify
polarization entanglement, which solves the problem above
perfectly \cite{polarization}, and called it Simon-Pan protocol.
However, in order to improve the fidelity of the entangled pairs
kept more with the PBS protocol \cite{panjw}, the two parties
should exploit quantum nondemolition (QND) measurement to
determine whether there are photons after the PBS or not, which
can not be accomplished only with PBS. Moreover, photon number
detectors should be used to distinguish the two-photon cases from
the cases with four photons in the same modes such as two photons
in the upper modes of  both Alice's and Bob's location. This task
can not be accomplished simply with linear optical elements.

Cross-Kerr nonlinearity provides a good tool to construct
nondestructive quantum nondemolition detectors  "which have the
potential available of being able to condition the evolution of
our system but without necessarily destroying the single photons"
\cite{QND,nl}. QND with a cross-Kerr medium and a coherent state
can be used for checking the parity of the polarizations of two
photons \cite{QND}, operating as a controlled-not (CNOT) gate
\cite{QND}, and analyzing the Bell states \cite{bellmeasuring}.
The Hamiltonian of a cross-Kerr nonlinear medium can be described
by the form as follows:
\begin{eqnarray}
H_{QND}=\hbar\chi\hat{n}_{a}\hat{n}_{c}
\end{eqnarray}
where $\hat{n}_{a}$ ($\hat{n}_{c}$) denotes the number operator
for mode a (c) and $\hbar\chi$ is the coupling strength of the
nonlinearity, which is decided by the property of material. For
example, for a signal photon state
$|\varphi\rangle=a|0\rangle+b|1\rangle$ and a coherent state
$|\alpha\rangle$, the cross-Kerr interaction causes the combined
system composed of a single photon and a coherent state to evolve
as \cite{QND}
\begin{eqnarray}
U_{ck}|\varphi\rangle|\alpha\rangle &=& e^{iH_{QND}t/\hbar}(a|0\rangle+b|1\rangle)|\alpha\rangle \nonumber\\
&=& a|0\rangle|\alpha\rangle+b|1\rangle|\alpha e^{i\theta}\rangle.
\end{eqnarray}
We note that $|0\rangle$ and $|1\rangle$ are not the polarization
of the photons, but the number of the photons. $|n\rangle$ is also
called the Fock state which means the state contains $n$ photons.
Now one can see that the signal photon state is unaffected by the
interaction, but the coherent state makes a phase shift of
$\theta$. Here $\theta=\chi t$ and $t$ is the interaction time.
The phase shift is directly proportional to the number of photons.
This is the main principle of the cross-Kerr nonlinearity
\cite{QND}. In 2005, Song \emph{et al.} \cite{song} presented a
protocol for entanglement purification using cross-Kerr
nonlinearity to complete parity check. It works for the original
entanglement purification model proposed by  Bennett et al.
\cite{C. H. Benneet}. The biggest advantage of their protocol is
that its successful probability can be nearly enhanced to a
quantity twice as large as that of PBS protocol \cite{panjw}. The
drawback of this protocol is the same as that in PBS protocol
\cite{panjw}. That is, it requires that the two parties of quantum
communication should be in possession of two ideal single-pair
entangled sources. Considering the currently available source of
entangled photons, this protocol becomes useless. Also, it cannot
get perfectly entangled photon pairs purified when the two parties
get a nonzero phase shift with $X$ homodyne measurements on their
coherent states, which takes place with a probability of the same
order of magnitude for the case where Alice and Bob both get the
phase shift $0$. Moreover, it can not complete the iteration of
the purification steps efficiently for improving the fidelity of
the entangled photons more.

In this paper, we present a way for  entanglement purification
based on two PDC sources with cross-Kerr nonlinearities. The task
of entanglement purification can be completed with two steps.
First, we provide a primary entanglement purification protocol for
PDC sources with QND detectors by transferring the spatial
entanglement of photon pairs to their polarization. In this
protocol, the QND detectors act as not only the role of CNot gates
but also that of photon number detectors, which provides a good
way for the next process to purify the entanglement of the photon
pairs more as they make the photon pairs equivalent to those
coming from two ideal sources.  In the second process for
entanglement purification, new QND detectors are designed to act
as the role of CNOT gates. This protocol has the advantage of high
yield and it requires neither CNOT gates based on linear optics
nor sophisticated single-photon detectors, which makes it more
convenient in practical applications.

\section{entanglement purification based on PDC sources}
\label{pep}

\subsection{ The principle of primary entanglement purification based on bit-flipping
errors with QND}

The principle of our entanglement purification protocol is shown
in Fig.1. The PDC sources can produce polarization and spatial
entanglement naturally \cite{polarization}. A pump pulse of
ultraviolet light passes through a beta barium borate (BBO)
crystal and produces correlated pairs of photons into the modes
$a_{1}$ and $b_{1}$. Then it is reflected and traverses the
crystal a second time, and produces correlated pairs of photons
into the modes $a_{2}$ and $b_{2}$. The Hamiltonian can be
approximately described as
\begin{eqnarray}
H_{PDC} &=& \gamma[(a^{+}_{1H}b^{+}_{1H}+a^{+}_{1V}b^{+}_{1V}) \nonumber\\
&+& r e^{i\phi}(a^{+}_{2H}b^{+}_{2H}+a^{+}_{2V}b^{+}_{2V})]
     + H.c,
\end{eqnarray}
where $H$ and $V$ in subscripts present  horizontal and vertical
polarization, $r$ denotes the relative probability of emission of
photons into the lower modes compared to the upper modes, and
$\phi$ is the phase between these two possibilities
\cite{polarization}. The same as the Simon-Pan protocol
\cite{polarization}, in a simple case we assume $r=1$ and
$\phi=0$. So the single-pair state can be described by
$(a^{+}_{1H}b^{+}_{1H}+a^{+}_{1V}b^{+}_{1V}+a^{+}_{2H}b^{+}_{2H}+a^{+}_{2V}b^{+}_{2V})|0\rangle$.
It  also can be written as
$(|a_{1}\rangle|b_{1}\rangle+|a_{2}\rangle|b_{2}\rangle)(|H_{a}\rangle|H_{b}\rangle
+ |V_{a}\rangle|V_{b}\rangle)$. The four-photon state produced by
this PDC source also can be written as
$(a^{+}_{1H}b^{+}_{1H}+a^{+}_{1V}b^{+}_{1V}+a^{+}_{2H}b^{+}_{2H}+a^{+}_{2V}b^{+}_{2V})^2|0\rangle$
and discussed in the same way.

\begin{figure}[!ht]
\begin{center}
\includegraphics[width=7cm,angle=0]{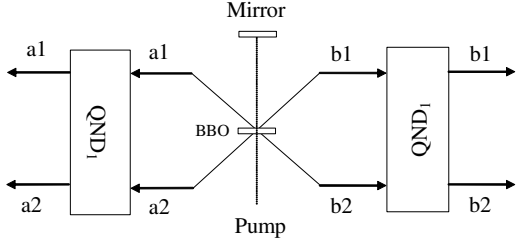}
\caption{The new entanglement purification protocol that uses  new
QND (QND$_1$) detectors and two parametric down-conversion
sources. The difference between this protocol and Simon-Pan
protocol \cite{polarization} is that we replace the two PBSs in
the latter with two QND$_1$ detectors. The PDC sources, which
produce two photons each into modes a1 and b1 and no photons into
modes a2 and b2 or vice versa, can be purified by  both Alice and
Bob selecting the same phase shifts. }
\end{center}
\end{figure}
\begin{figure}[!ht]
\begin{center}
\includegraphics[width=7cm,angle=0]{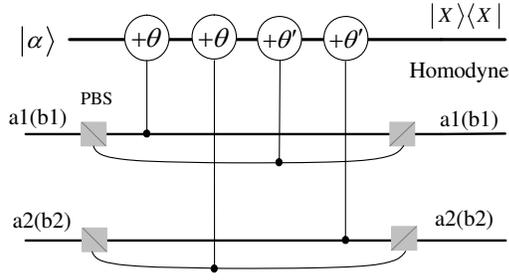}
\caption{Schematic diagram showing the principle of our new
nondestructive quantum nondemolition (QND$_1$) detectors. Several
cross-Kerr nonlinearities and a coherent laser probe beam
$|\alpha\rangle$ are used in our protocol. This QND can transform
spatial entanglement into polarization entanglement. It  can also
distinguish superpositions and mixtures of the states $|HH\rangle$
and $|VV\rangle$ from $|VH\rangle$ and $|HV\rangle$. It acts as
not only the role of CNot gates but also of photon number
detectors here.}
\end{center}
\end{figure}
\begin{figure}[!ht]
\begin{center}
\includegraphics[width=4cm,angle=0]{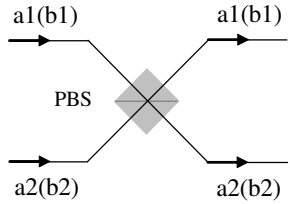}
\caption{Schematic diagram showing the principle of a coupler.}
\end{center}
\end{figure}

After receiving the signals, the user Alice (Bob) lets them pass
through QND$_1$ detectors whose principle is shown in Fig.2. For a
two-photon state without suffering from decoherence (including
bit-flipping and phase-flipping)
$(a^{+}_{1H}b^{+}_{1H}+a^{+}_{1V}b^{+}_{1V}+a^{+}_{2H}b^{+}_{2H}+a^{+}_{2V}b^{+}_{2V})|0\rangle$,
the two parties Alice and Bob will get the same phase shifts on
their coherent states as QND$_1$ detectors evolve the combined
system to
\begin{eqnarray}
&\rightarrow &
(a^{+}_{1H}b^{+}_{1H}+a^{+}_{2V}b^{+}_{2V})|0\rangle |\alpha
e^{i\theta}\rangle_a |\alpha e^{i\theta}\rangle_b \nonumber\\
&& +(a^{+}_{1V}b^{+}_{1V}+a^{+}_{2H}b^{+}_{2H})|0\rangle |\alpha
e^{i\theta'}\rangle_a |\alpha e^{i\theta'}\rangle_b,
\end{eqnarray}
where $\theta \neq \theta' \oplus 2 \pi$. If Alice and Bob get the
same results with an $X$ homodyne measurement ($\theta$ or
$\theta'$), they retain the pair and perform no local unitary
operations on their photons but link the photons with couplers
shown in Fig.3. If both Alice and Bob get the phase shift $\theta$
($\theta'$), their photon pair in the state
$(a^{+}_{H}b^{+}_{H}+a^{+}_{V}b^{+}_{V})|0\rangle$ will appear at
the lower output modes $a_2b_2$ (upper modes $a_1b_1$) of the
couplers. If a bit-flipping error takes place, i.e., the state of
the pair becoming
$(|a_{1}\rangle|b_{1}\rangle+|a_{2}\rangle|b_{2}\rangle)(|V_{a}\rangle|H_{b}\rangle
+ |H_{a}\rangle|V_{b}\rangle)$,  Alice and Bob will get two
different results with their homodyne measurements on their
coherent states $|\alpha\rangle$ as QND$_1$ detectors evolve the
combined system to
\begin{eqnarray}
&\rightarrow &
(a^{+}_{1V}b^{+}_{1H}+a^{+}_{2H}b^{+}_{2V})|0\rangle |\alpha
e^{i\theta'}\rangle_a |\alpha e^{i\theta}\rangle_b \nonumber\\
&& +(a^{+}_{1H}b^{+}_{1V}+a^{+}_{2V}b^{+}_{2H})|0\rangle |\alpha
e^{i\theta}\rangle_a |\alpha e^{i\theta'}\rangle_b.
\end{eqnarray}
One will get the result $\theta$ and the other $\theta'$.
Therefore, by performing a bit-flipping operation
$\sigma_x=|H\rangle\langle V| + |V\rangle\langle H|$, Alice and
Bob can get rid of all bit-flip errors and obtain their
uncorrupted pairs
$(a^{+}_{H}b^{+}_{H}+a^{+}_{V}b^{+}_{V})|0\rangle$ by coupling the
two spatial modes with their couplers.

Certainly, a phase-flipping error can not be directly purified in
this way. However, as pointed out by others \cite{C. H. Benneet,C.
H. Benneet2,D. Deutsch,panjw}, a phase-flipping error can be
transformed into a bit-flipping error using a bilateral local
operation. If a bit-flipping error purification has been
successfully solved, phase-flipping errors also can be solved
perfectly. In this way the two parties can purify a general mixed
state. We only discuss the case with bit-flipping errors below.

For the four-photon state
$(a^{+}_{1H}b^{+}_{1H}+a^{+}_{1V}b^{+}_{1V}+a^{+}_{2H}b^{+}_{2H}+a^{+}_{2V}b^{+}_{2V})^2|0\rangle$
which has the same order of magnitude of probability as the
two-photon state
$(a^{+}_{1H}b^{+}_{1H}+a^{+}_{1V}b^{+}_{1V}+a^{+}_{2H}b^{+}_{2H}+a^{+}_{2V}b^{+}_{2V})|0\rangle$,
if it does not suffer  from decoherence, the QND$_1$ detectors
evolve the combined system to
\begin{eqnarray}
&\rightarrow & (a^{+}_{1H}b^{+}_{1H} + a^+_{2V}b^+_{2V})^2
|0\rangle |\alpha e^{i2\theta}\rangle_a |\alpha
e^{i2\theta}\rangle_b  \nonumber\\
&& +  (a^{+}_{1V}b^{+}_{1V} + a^+_{2H}b^+_{2H})^2 |0\rangle
|\alpha
e^{i2\theta'}\rangle_a |\alpha e^{i2\theta'}\rangle_b \nonumber\\
 &&
+2(a^+_{1H}b^+_{1H}  +  a^+_{2V}b^+_{2V}) (a^+_{2H}b^+_{2H}
\nonumber\\
&&+ a^+_{1V}b^+_{1V})|0\rangle |\alpha
e^{i(\theta+\theta')}\rangle_a |\alpha
e^{i(\theta+\theta')}\rangle_b.
\end{eqnarray}
Similar to the case with the two-photon state, Alice and Bob will
get the same phase shifts with their homodyne measurements on
their coherent states. That is, they both get $2\theta$,
$2\theta'$, or $\theta+\theta'$ which corresponds to the
four-photon state $(a^{+}_{1H}b^{+}_{1H} + a^+_{2V}b^+_{2V})^2
|0\rangle$, $(a^{+}_{1V}b^{+}_{1V} + a^+_{2H}b^+_{2H})^2
|0\rangle$, or $(a^+_{1H}b^+_{1H}  +  a^+_{2V}b^+_{2V})
(a^+_{2H}b^+_{2H} + a^+_{1V}b^+_{1V})|0\rangle$, respectively. The
state $(a^+_{1H}b^+_{1H}  +  a^+_{2V}b^+_{2V}) (a^+_{2H}b^+_{2H} +
a^+_{1V}b^+_{1V})|0\rangle$ represents the case that one pair
appears at the upper modes and the other at the lower modes after
the couplers, and  both in the desired state
$(a^{+}_{H}b^{+}_{H}+a^{+}_{V}b^{+}_{V})|0\rangle$. The state
$(a^{+}_{1H}b^{+}_{1H} + a^+_{2V}b^+_{2V})^2 |0\rangle$
($(a^{+}_{1V}b^{+}_{1V} + a^+_{2H}b^+_{2H})^2 |0\rangle$) denotes
that the two pairs both appear at the lower (upper) modes after
the couplers. That is, the QND$_1$ detectors can pick up the state
wanted from others with the spatial entanglement resource.

If a bit-flipping error takes place on one of the two photon pairs
in the four-photon state, i.e., the state of the two photon pairs
becoming
$(a^{+}_{1H}b^{+}_{1H}+a^{+}_{1V}b^{+}_{1V}+a^{+}_{2H}b^{+}_{2H}+a^{+}_{2V}b^{+}_{2V})
(a^{+}_{1V}b^{+}_{1H}+a^{+}_{1H}b^{+}_{1V}+a^{+}_{2V}b^{+}_{2H}+a^{+}_{2H}b^{+}_{2V})|0\rangle$,
the QND$_1$ detectors evolve the combined system to
\begin{widetext}
\begin{center}
\begin{eqnarray}
&\rightarrow & (a^{+}_{1H}b^{+}_{1H} +
a^+_{2V}b^+_{2V})(a^+_{1V}b^+_{1H} + a^+_{2H}b^+_{2V}) |0\rangle
|\alpha
e^{i(\theta + \theta')}\rangle_a |\alpha e^{i2\theta}\rangle_b \nonumber\\
 &&
+  (a^{+}_{1H}b^{+}_{1H} + a^+_{2V}b^+_{2V})(a^+_{1H}b^+_{1V} +
a^+_{2V}b^+_{2H}) |0\rangle |\alpha
e^{i2\theta}\rangle_a |\alpha e^{i(\theta + \theta')}\rangle_b \nonumber\\
 &&
+  (a^{+}_{1V}b^{+}_{1V} + a^+_{2H}b^+_{2H})(a^+_{1V}b^+_{1H} +
a^+_{2H}b^+_{2V}) |0\rangle |\alpha
e^{i2\theta'}\rangle_a |\alpha e^{i(\theta + \theta')}\rangle_b \nonumber\\
 &&
+  (a^{+}_{1V}b^{+}_{1V} +a^+_{2H}b^+_{2H})(a^+_{1H}b^+_{1V} +
a^+_{2V}b^+_{2H}) |0\rangle |\alpha e^{i(\theta +
\theta')}\rangle_a |\alpha e^{i2\theta'}\rangle_b.
\end{eqnarray}
\end{center}
\end{widetext}
That is, Alice and Bob can not get the same phase shifts with
their homodyne measurements on their coherent states. After the
couplers, one of the two parties gets two photons coming from two
modes but the other gets two photons from only one mode, which
makes Alice and Bob have no ability to get the uncorrupted state
$(a^{+}_{H}b^{+}_{H}+a^{+}_{V}b^{+}_{V})|0\rangle$ perfectly.
Alice and Bob discard all these instances, the same as the
Simon-Pan protocol \cite{polarization}.

If  bit-flipping errors take place on both the two photon pairs in
the four-photon state, i.e., the state of the two pairs becoming
$(a^{+}_{1V}b^{+}_{1H}+a^{+}_{1H}b^{+}_{1V}+a^{+}_{2V}b^{+}_{2H}+a^{+}_{2H}b^{+}_{2V})^2
|0\rangle$, QND$_1$ detectors evolve the combined system to
\begin{eqnarray}
&\rightarrow & (a^{+}_{1H}b^{+}_{1V} + a^+_{2V}b^+_{2H})^2
|0\rangle |\alpha e^{i2\theta}\rangle_a |\alpha
e^{i2\theta'}\rangle_b  \nonumber\\
&& +  (a^{+}_{1V}b^{+}_{1H} + a^+_{2H}b^+_{2V})^2 |0\rangle
|\alpha
e^{i2\theta'}\rangle_a |\alpha e^{i2\theta}\rangle_b \nonumber\\
 &&
+2(a^+_{1V}b^+_{1H}  +  a^+_{2H}b^+_{2V}) (a^+_{1H}b^+_{1V}
\nonumber\\
&&+ a^+_{2V}b^+_{2H})|0\rangle |\alpha
e^{i(\theta+\theta')}\rangle_a |\alpha
e^{i(\theta+\theta')}\rangle_b.
\end{eqnarray}
Alice and Bob should discard the instances for which one gets the
phase shift $2\theta$ and the other gets $2\theta'$ as the two
photon pairs appear at the same mode simultaneously. When both get
the phase shift $\theta + \theta'$, they will keep these unwanted
photon pairs  as they can not distinguish the corrupted photon
pairs from the uncorrupted ones in this way.

In a practical application, the spatial entanglement in this
two-photon state is completely transformed into the polarization
entanglement in the process for eliminating the bit-flipping
errors. They can not be used again for correcting the
phase-flipping errors directly. It requires the two parties to
exploit another purification protocol to solve this problem. It is
valuable to point out that the QND$_1$ detectors act as not only a
nondestructive measurement tool but also a tool for distinguishing
the number of photons. This tool is very useful for the next
purification to improve the fidelity of the pairs more.

\subsection{The fidelity of the photon pairs}

Now let us pay our attention to the fidelity of the photon pairs.
Suppose the probabilities of one photon pair and two photon pairs
produced by  PDC sources are $p_1$ and $p_2$, respectively.
Suppose the probability of bit-flipping arisen from the quantum
channel such as fibers is $1-F_0$, which means the original
fidelity of the photon pairs controlled by the two users Alice and
Bob is $F_0$.

In our primary entanglement purification protocol, the two-photon
states are all be kept, which takes place with the probability
$p_1$, and their fidelity is 1 after Alice performs a bit-flipping
operation or not. For the four-photon states, Alice and Bob only
keep the cases where each mode has one and only one photon, which
takes place with the probability of $\frac{1}{2}p_2[F_0^2 +
(1-F_0)^2]$. After the primary entanglement purification, the
fidelity of the photon pairs kept becomes
\begin{eqnarray}
F_1=\frac{p_1+\frac{1}{2}p_2F_0^2}{p_1+\frac{1}{2}p_2[F_0^2+(1-F_0)^2]}.
\end{eqnarray}

\section{entanglement purification based on ideal sources}
\label{epis}

After the primary entanglement purification based on PDC sources
in Sec. \ref{pep}, the photon pairs kept are equivalent to those
coming from two ideal sources as the QND$_1$ can distinguish the
two-photon states from the four-photon states. Moreover, it shows
there are useful photon pairs or not clearly for the two users. In
this time, Alice and Bob can exploit the entanglement purification
protocols with CNOT gates such as those in Refs. \cite{C. H.
Benneet,D. Deutsch} or the PBS protocol proposed by Pan et al.
\cite{panjw} to improve the fidelity of the photon pairs more. At
present, a CNOT gate with single photons is far beyond what is
experimentally feasible. The PBS protocol requires sophisticated
single-photon detectors and its yield of photon pairs purified is
only half of that with CNOT gates. The protocol in Ref.
\cite{song} with  QND detectors designed by Nemoto and Munro
\cite{QND} can also be used to purify less entangled pairs with
$X$ quadrature measurements \cite{xm} in a nearly deterministic
way as the two users Alice and Bob should ensure that the states
$|\alpha e^{\pm i \theta}\rangle$ can not be distinguished.

In this section, we will present a different entanglement
purification protocol for ideal sources in a completely
deterministic way without CNOT gates and sophisticated
single-photon detractors. It has the same yield of photon pairs
purified as those \cite{C. H. Benneet,D. Deutsch} with CNOT gates,
double that of the PBS protocol \cite{panjw}.

\begin{figure}[!ht]
\begin{center}
\includegraphics[width=6cm,angle=0]{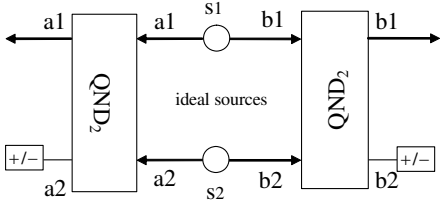}
\caption{The new entanglement purification protocol that uses  new
QND (QND$_2$) detectors and two ideal sources (S1 and S2). It can
be used as the second purification process for improving the
fidelity of photon pairs from PDC sources more. }
\end{center}
\end{figure}
\begin{figure}[!ht]
\begin{center}
\includegraphics[width=6cm,angle=0]{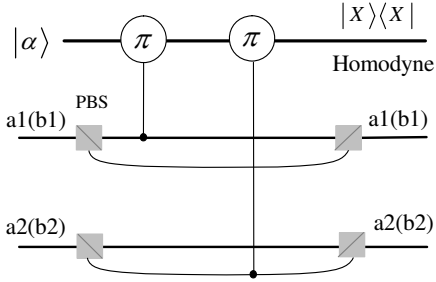}
\caption{Schematic diagram showing the principle of our new
nondestructive quantum nondemolition detectors (QND$_2$). Each
QND$_2$ is composed of two same cross-Kerr nonlinearities with the
phase shift $\theta=\chi t=\pi$, four PBSs and a coherent laser
probe beam $|\alpha\rangle$.}
\end{center}
\end{figure}

Suppose the photon pairs after the primary entanglement
purification are in the mixed state $\rho_{ab}$, the same as that
in Ref.\cite{panjw}, described as follows:
\begin{eqnarray}
\rho_{ab}=F|\Phi^{+}\rangle_{ab}\langle\Phi^{+}|+(1-F)|\Psi^{+}\rangle_{ab}\langle\Psi^{+}|,
\end{eqnarray}
where
$|\Phi^{+}\rangle_{ab}=\frac{1}{\sqrt{2}}(|H\rangle_{a}|H\rangle_{b}+|V\rangle_{a}|V\rangle_{b})$
and
$|\Psi^{+}_{ab}\rangle=\frac{1}{\sqrt{2}}(|H\rangle_{a}|V\rangle_{b}+|V\rangle_{a}|H\rangle_{b})$.
$F$ ($>\frac{1}{2}$) is the fidelity of the  state, i.e.,
$F=\langle\Phi^{+}|\rho_{ab}|\Phi^{+}\rangle$. The two photons in
the state $|\Phi^{+}\rangle_{ab}$  have the equal polarizations
and those in the state $|\Psi^{+}_{ab}\rangle$ have the opposite
polarizations in the rectangle basis $\{\vert H\rangle, \vert
V\rangle\}$. The two pairs can be seen as the mixture of four
states, i.e.,
$|\Phi^{+}\rangle_{a1b1}\cdot|\Phi^{+}\rangle_{a2b2}$ with a
probability of $F^{2}$, both
$|\Phi^{+}\rangle_{a1b1}\cdot|\Psi^{+}\rangle_{a2b2}$ and
$|\Psi^{+}\rangle_{a1b1}\cdot|\Phi^{+}\rangle_{a2b2}$ with an
equal probability of $F(1-F)$, and
$|\Psi^{+}\rangle_{a1b1}\cdot|\Psi^{+}\rangle_{a2b2}$ with a
probability of $(1-F)^{2}$.

The principle of our entanglement purification protocol based on
two ideal sources is shown in Fig.4. It is composed of two QND
detectors (QND$_2$) and two measurements with the diagonal basis
$\{|\pm\rangle =\frac{1}{\sqrt{2}}(|H\rangle \pm |V\rangle)\}$.
The principle of QND$_2$ is shown in Fig.5. The two same
cross-Kerr nonlinearities provide the same phase shift
$\theta=\chi' t=\pi$.

Let us first consider the state
$|\Phi^{+}\rangle_{a1b1}\cdot|\Phi^{+}\rangle_{a2b2}$.
\begin{eqnarray}
&&|\Phi^{+}\rangle_{a1b1}\cdot|\Phi^{+}\rangle_{a2b2} =
\frac{1}{\sqrt{2}}(|H\rangle_{a1}|H\rangle_{b1}+
|V\rangle_{a1}|V\rangle_{b1}) \nonumber\\
&&\;\;\;\;\;\;\;\;\;\;\;\;\;\;\;\;\;\;\;\;\;\;\;\;\;\;\;\;\;\;
\otimes
\frac{1}{\sqrt{2}}(|H\rangle_{a2}|H\rangle_{b2}+|V\rangle_{a2}|V\rangle_{b2})\nonumber\\
&&=
\frac{1}{2}(|H\rangle_{a1}|H\rangle_{b1}|H\rangle_{a2}|H\rangle_{b2}+
|H\rangle_{a1}|H\rangle_{b1}|V\rangle_{a2}|V\rangle_{b2}  \nonumber\\
&&
+|V\rangle_{a1}|V\rangle_{b1}|H\rangle_{a2}|H\rangle_{b2}+|V\rangle_{a1}|V\rangle_{b1}|V\rangle_{a2}|V\rangle_{b2}).\nonumber\\
\end{eqnarray}
QND$_2$ detectors evolve the combined system composed of four
photons and two coherent states to
\begin{eqnarray}
&\rightarrow & \frac{1}{2}\{(|H\rangle_{a1} |H\rangle_{b1}
|H\rangle_{a2} |H\rangle_{b2} \nonumber\\
&& + |V\rangle_{a1} |V\rangle_{b1} |V\rangle_{a2}
|V\rangle_{b2})|\alpha e^{i\pi}\rangle_a |\alpha
e^{i\pi}\rangle_b \nonumber\\
 && + |H\rangle_{a1} |H\rangle_{b1} |V\rangle_{a2} |V\rangle_{b2} |\alpha e^{i2\pi}\rangle_a
|\alpha e^{i2\pi}\rangle_b  \nonumber\\
&& + |V\rangle_{a1} |V\rangle_{b1} |H\rangle_{a2} |H\rangle_{b2}
|\alpha \rangle_a |\alpha \rangle_b\}.
\end{eqnarray}
When both Alice and Bob get the phase shift $\pi$ with their
homodyne measurements on their coherent states, the two photon
pairs project to the state $(|H\rangle_{a1} |H\rangle_{b1}
|H\rangle_{a2} |H\rangle_{b2} + |V\rangle_{a1} |V\rangle_{b1}
|V\rangle_{a2} |V\rangle_{b2})$. When they both get the phase
shift $0$ ($2\pi$ is just the phase shift $0$ for the coherent
states), they get the state $(|H\rangle_{a1} |H\rangle_{b1}
|V\rangle_{a2} |V\rangle_{b2} + |V\rangle_{a1} |V\rangle_{b1}
|H\rangle_{a2} |H\rangle_{b2})$ and they can obtain the state
$(|H\rangle_{a1} |H\rangle_{b1} |H\rangle_{a2} |H\rangle_{b2} +
|V\rangle_{a1} |V\rangle_{b1} |V\rangle_{a2} |V\rangle_{b2})$ by
performing a bit-flipping operation $\sigma_x=|H\rangle\langle V|
+ |V\rangle\langle H|$ on their first photons $a1$ and $b1$. With
the same way as in Ref.\cite{panjw} Alice and Bob can make the
photon pair in the state $|\Phi^+\rangle_{ab}$. In detail, Alice
and Bob first take a measurement with the diagonal basis on their
second photons $a_2$ and $b_2$. When they both get the results
$\vert +\rangle$ (or $\vert -\rangle$), the photon pairs $a_1b_1$
are projected to the state $|\Phi^+\rangle_{ab}$. When one gets
the result $\vert +\rangle$ and the other gets $\vert -\rangle$,
they can obtain the state $|\Phi^+\rangle_{ab}$ by performing the
phase-flipping $\sigma_z=|H\rangle\langle H| - |V\rangle\langle
V|$ on the photon $a_1$.

For the cross-combinations
$|\Phi^{+}\rangle_{a1b1}\cdot|\Psi^{+}\rangle_{a2b2}$ and
$|\Psi^{+}\rangle_{a1b1}\cdot|\Phi^{+}\rangle_{a2b2}$, the QND$_2$
detectors will evolve the combined system to the state in which
Alice and Bob can not get the same phase shift with their homodyne
measurements on their coherent states. In detail,
$|\Phi^{+}\rangle_{a1b1}\cdot|\Psi^{+}\rangle_{a2b2}$ will be
evolved to
\begin{eqnarray}
&\rightarrow & \frac{1}{2}\{(|H\rangle_{a1} |H\rangle_{b1}
|V\rangle_{a2} |H\rangle_{b2}  \nonumber\\
&&+ |V\rangle_{a1} |V\rangle_{b1} |H\rangle_{a2}
|V\rangle_{b2})|\alpha \rangle_a |\alpha
e^{i\pi}\rangle_b \nonumber\\
 && + |H\rangle_{a1} |H\rangle_{b1} |H\rangle_{a2} |V\rangle_{b2} |\alpha e^{i\pi}\rangle_a
|\alpha e^{i2\pi}\rangle_b  \nonumber\\
&& + |V\rangle_{a1} |V\rangle_{b1} |V\rangle_{a2} |H\rangle_{b2}
|\alpha e^{i\pi}\rangle_a |\alpha \rangle_b\},
\end{eqnarray}
and $|\Psi^{+}\rangle_{a1b1}\cdot|\Phi^{+}\rangle_{a2b2}$ will be
evolved to
\begin{eqnarray}
&\rightarrow & \frac{1}{2}\{(|V\rangle_{a1} |H\rangle_{b1}
|H\rangle_{a2} |H\rangle_{b2}\nonumber\\
&& + |H\rangle_{a1} |V\rangle_{b1} |V\rangle_{a2}
|V\rangle_{b2})|\alpha \rangle_a |\alpha
e^{i\pi}\rangle_b \nonumber\\
 && + |V\rangle_{a1} |H\rangle_{b1} |V\rangle_{a2} |V\rangle_{b2} |\alpha e^{i\pi}\rangle_a
|\alpha e^{i2\pi}\rangle_b \nonumber\\
&& + |H\rangle_{a1} |V\rangle_{b1} |H\rangle_{a2} |H\rangle_{b2}
|\alpha e^{i\pi}\rangle_a |\alpha \rangle_b\}.
\end{eqnarray}
When Alice gets the phase shift $0$ and Bob gets $\pi$, their two
photon pairs $a_1b_1$ and $a_2b_2$ are in the state
$(|H\rangle_{a1} |H\rangle_{b1} |V\rangle_{a2} |H\rangle_{b2} +
|V\rangle_{a1} |V\rangle_{b1} |H\rangle_{a2} |V\rangle_{b2})$ or
$(|V\rangle_{a1} |H\rangle_{b1} |H\rangle_{a2} |H\rangle_{b2} +
|H\rangle_{a1} |V\rangle_{b1} |V\rangle_{a2} |V\rangle_{b2})$ with
the same probability. In this time, Alice and Bob can not
determine in which pair takes place a bit-flipping error. For
improving the fidelity of the photon pairs kept, Alice and Bob
should discard both these photon pairs, the same as that in the
protocol with CNOT gates \cite{C. H. Benneet,D. Deutsch}. When
Alice gets the phase shift $\pi$ and Bob gets $0$, they should
also discard their two photon pairs.

For the state
$|\Psi^{+}\rangle_{a1b1}\cdot|\Psi^{+}\rangle_{a2b2}$, QND$_2$
detectors evolve the combined system to
\begin{eqnarray}
&\rightarrow & \frac{1}{2}\{(|V\rangle_{a1} |H\rangle_{b1}
|V\rangle_{a2} |H\rangle_{b2}  \nonumber\\
&& + |H\rangle_{a1} |V\rangle_{b1} |H\rangle_{a2}
|V\rangle_{b2})|\alpha e^{i\pi}\rangle_a |\alpha
e^{i\pi}\rangle_b \nonumber\\
 && + (|V\rangle_{a1} |H\rangle_{b1} |H\rangle_{a2} |V\rangle_{b2}  \nonumber\\
 && + |H\rangle_{a1} |V\rangle_{b1} |V\rangle_{a2}
|H\rangle_{b2}) |\alpha \rangle_a |\alpha \rangle_b\}.
\end{eqnarray}
When Alice and Bob both get the phase shift $\pi$, their two
photon pairs are in the state $(|V\rangle_{a1} |H\rangle_{b1}
|V\rangle_{a2} |H\rangle_{b2} + |H\rangle_{a1} |V\rangle_{b1}
|H\rangle_{a2} |V\rangle_{b2})$. After Alice and Bob perform a
measurement with the diagonal basis on their second photons $a_2$
and $b_2$, the first photon pair $a_1b_1$ projects to the state
$\vert \Psi^+\rangle_{ab}$ when they both obtain the outcome
$|+\rangle$ (or $|-\rangle$); otherwise Alice and Bob will make
the pair $a_1b_1$ in this state by performing a phase-flipping
operation $\sigma_z$. When Alice and Bob both get the phase shift
$0$, their two photon pairs are in the state $(|V\rangle_{a1}
|H\rangle_{b1} |H\rangle_{a2} |V\rangle_{b2}  + |H\rangle_{a1}
|V\rangle_{b1} |V\rangle_{a2} |H\rangle_{b2})$. With the same
operations as those in the case where both  photon pairs do not
contain errors, Alice and Bob will make their first photon pair in
the state $\vert \Psi^+\rangle_{ab}$. In other words, Alice and
Bob can not distinguish the two cases that contain no errors in
their two photon pairs or that both contain a bit-flipping error.
They keep those photon pairs for improving their fidelity in the
next round.

By postselection according to the phase shifts of the coherent
states, Alice and Bob only keep the first photon pair in the
instances where they get the same phase shifts. After this
purification process, the new fidelity of the photon pairs kept
becomes
\begin{eqnarray}
F'=\frac{F^{2}}{F^{2}+(1-F)^{2}}.
\end{eqnarray}
We get the same fidelity as in the PBS protocol \cite{panjw}, but
the yield is double  that in the PBS protocol as Alice and Bob
will keep a photon pair when they get the same phase shift, no
matter what it is. In PBS protocol, Alice and Bob only keep the
instances that each mode has one and only one photon, which makes
its yield half  those with CNOT gates  \cite{C. H. Benneet,D.
Deutsch}. Moreover, Alice and Bob use the homodyne measurements on
their coherent states to replace the sophisticated single-photon
detectors in PBS protocol \cite{panjw}. This new entanglement
purification protocol can be used to improve the fidelity of
photon pairs more by iteration.

\section{discussion and summary}

In the primary entanglement purification protocol, Alice and Bob
can also use the QND$_3$, whose principle is shown in Fig.6, to
purify the photon pairs produced by two PDC sources if they can
control accurately the overlap time of the photons coming from the
upper mode and the lower mode. In essence, the two parties exploit
the cross-Kerr nonlinearities, instead of the sophisticated
single-photon detectors in the Simon-Pan protocol
\cite{polarization}, to complete the task of distinguishing the
photon numbers from their modes, without destroying the photons in
this time.

\begin{figure}[!ht]
\begin{center}
\includegraphics[width=6cm,angle=0]{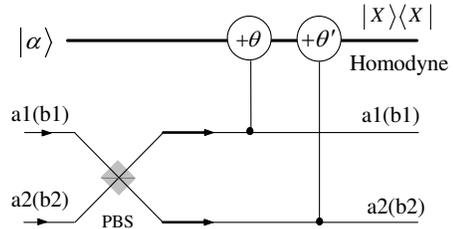}
\caption{Schematic diagram showing the principle of simple
nondestructive quantum nondemolition detectors (QND$_3$) for
purifying the photon pairs produced from two PDC sources. Each
QND$_3$ is composed of two different cross-Kerr nonlinearities
with the phase shift $\theta$ and $\theta'$, a PBS and a coherent
laser probe beam $|\alpha\rangle$.}
\end{center}
\end{figure}

For a two-photon state without suffering from decoherence
$(a^{+}_{1H}b^{+}_{1H}+a^{+}_{1V}b^{+}_{1V}+a^{+}_{2H}b^{+}_{2H}+a^{+}_{2V}b^{+}_{2V})|0\rangle$,
the two parties Alice and Bob will get the same phase shift on
their coherent states as QND$_3$ detectors evolve the combined
system to
\begin{eqnarray}
&\rightarrow &
(a^{+}_{1H}b^{+}_{1H}+a^{+}_{1V}b^{+}_{1V})|0\rangle |\alpha
e^{i\theta}\rangle_a |\alpha e^{i\theta}\rangle_b \nonumber\\
&& +(a^{+}_{2V}b^{+}_{2V}+a^{+}_{2H}b^{+}_{2H})|0\rangle |\alpha
e^{i\theta'}\rangle_a |\alpha e^{i\theta'}\rangle_b,
\end{eqnarray}
where $\theta \neq \theta' \oplus 2\pi$. If Alice and Bob get the
same results with an $X$ homodyne measurement ($\theta$ or
$\theta'$), they get a photon pair in the state
$(a^{+}_{H}b^{+}_{H}+a^{+}_{V}b^{+}_{V})|0\rangle$. The homodyne
measurement provides not only the information about the
polarization state of the photon pair but also their spatial
modes.  If a bit-flipping error takes place, i.e., the state of
the pair becoming
$(|a_{1}\rangle|b_{1}\rangle+|a_{2}\rangle|b_{2}\rangle)(|V_{a}\rangle|H_{b}\rangle
+ |H_{a}\rangle|V_{b}\rangle)$,  Alice and Bob will get two
different results with their homodyne measurements on their
coherent states $|\alpha\rangle$ as QND$_3$ detectors evolve the
combined system to
\begin{eqnarray}
&\rightarrow &
(a^{+}_{1V}b^{+}_{2H}+a^{+}_{1H}b^{+}_{2V})|0\rangle |\alpha
e^{i\theta}\rangle_a |\alpha e^{i\theta'}\rangle_b \nonumber\\
&& +(a^{+}_{2V}b^{+}_{1H}+a^{+}_{2H}b^{+}_{1V})|0\rangle |\alpha
e^{i\theta'}\rangle_a |\alpha e^{i\theta}\rangle_b.
\end{eqnarray}
One will get the result $\theta$ and the other $\theta'$. By
performing a bit-flipping operation $\sigma_x=|H\rangle\langle V|
+ |V\rangle\langle H|$ on one photon such as the photon controlled
by Alice, Alice and Bob can get rid of all bit-flip errors and
obtain their uncorrupted pair
$(a^{+}_{H}b^{+}_{H}+a^{+}_{V}b^{+}_{V})|0\rangle$.

For the four-photon state
$(a^{+}_{1H}b^{+}_{1H}+a^{+}_{1V}b^{+}_{1V}+a^{+}_{2H}b^{+}_{2H}+a^{+}_{2V}b^{+}_{2V})^2|0\rangle$,
the QND$_3$ detectors evolve the combined system to
\begin{eqnarray}
&\rightarrow & (a^{+}_{1V}b^{+}_{1V} + a^+_{1H}b^+_{1H})^2
|0\rangle |\alpha e^{i2\theta}\rangle_a |\alpha
e^{i2\theta}\rangle_b   \nonumber\\
&& +   (a^{+}_{2H}b^{+}_{2H} + a^+_{2V}b^+_{2V})^2 |0\rangle
|\alpha e^{i2\theta'}\rangle_a |\alpha
e^{i2\theta'}\rangle_b\nonumber\\
 &&
+2(a^+_{1H}b^+_{1H}  +  a^+_{1V}b^+_{1V}) (a^+_{2H}b^+_{2H}
\nonumber\\
&& + a^+_{2V}b^+_{2V})|0\rangle |\alpha
e^{i(\theta+\theta')}\rangle_a |\alpha
e^{i(\theta+\theta')}\rangle_b.
\end{eqnarray}
Alice and Bob only pick up the four-mode instances, i.e., they get
the same phase shift $\theta + \theta'$. If a bit-flipping error
takes place on one of the two photon pairs in the four-photon
state, i.e., the state of the two photon pairs becoming
$(a^{+}_{1H}b^{+}_{1H}+a^{+}_{1V}b^{+}_{1V}+a^{+}_{2H}b^{+}_{2H}+a^{+}_{2V}b^{+}_{2V})
(a^{+}_{1V}b^{+}_{1H}+a^{+}_{1H}b^{+}_{1V}+a^{+}_{2V}b^{+}_{2H}+a^{+}_{2H}b^{+}_{2V})|0\rangle$,
the four photons pass through only three ports \cite{polarization}
after PBS. In this way, Alice and Bob can not get the same phase
shift $\theta + \theta'$ when they measure their coherent states.
When bit-flipping errors take place on both photon pairs in the
four-photon state, Alice and Bob will get the same phase shift
$\theta + \theta'$ as QND$_3$ detectors evolve the combined system
from the state
$(a^{+}_{1V}b^{+}_{1H}+a^{+}_{1H}b^{+}_{1V}+a^{+}_{2V}b^{+}_{2H}+a^{+}_{2H}b^{+}_{2V})^2
|0\rangle |\alpha\rangle_a |\alpha\rangle_b$ to
\begin{eqnarray}
&\rightarrow & (a^{+}_{2H}b^{+}_{1V} + a^+_{2V}b^+_{1H})^2
|0\rangle |\alpha e^{i2\theta'}\rangle_a |\alpha
e^{i2\theta}\rangle_b  \nonumber\\
&& +  (a^{+}_{1V}b^{+}_{2H} + a^+_{1H}b^+_{2V})^2 |0\rangle
|\alpha
e^{i2\theta}\rangle_a |\alpha e^{i2\theta'}\rangle_b \nonumber\\
 &&
+2(a^+_{1V}b^+_{2H}  +  a^+_{1H}b^+_{2V}) (a^+_{2H}b^+_{1V}
\nonumber\\
&& + a^+_{2V}b^+_{1H})|0\rangle |\alpha
e^{i(\theta+\theta')}\rangle_a |\alpha
e^{i(\theta+\theta')}\rangle_b.
\end{eqnarray}

\begin{figure}[!ht]
\begin{center}
\includegraphics[width=6cm,angle=0]{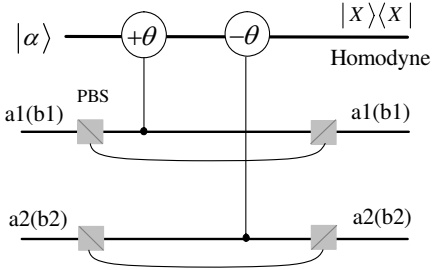}
\caption{The principle of the nondestructive quantum nondemolition
detectors (QND$_4$) designed by Nemoto and Munro for parity check
\cite{QND}. Two different cross-Kerr nonlinearities have the phase
shift $\theta$ and $-\theta$, respectively.}
\end{center}
\end{figure}

From the discussion above, one can see that QND$_3$ detectors act
as the same role as QND$_1$ if the two parties can  control
accurately the overlap time of the photons coming from the upper
mode ($a_1(b_1)$) and the lower mode ($a_2(b_2)$). In experiment,
each party need only use two different cross-Kerr nonlinearities,
not four, which may  make it more convenient than the QND$_1$
detectors.

In this entanglement purification based on ideal sources in Sec.
\ref{epis} we do not exploit the QND detector designed by Nemoto
and Munro \cite{QND} (namely QND$_4$ shown in Fig.7) as the
QND$_2$ is more efficient than the latter. If Alice and Bob
exploit QND$_4$ detectors to purify the two photon pairs produced
by two ideal sources, they should perform a sophisticated $X$
quadrature measurement in which the states $|\alpha e^{\pm
\theta}\rangle$ can not be distinguished \cite{QND,xm} as QND$_4$
evolve the combined system from the state $(|H\rangle_{a_1}
|H\rangle_{b_1} |H\rangle_{a_2} |H\rangle_{b_2} + |V\rangle_{a_1}
|V\rangle_{b_1} |V\rangle_{a_2} |V\rangle_{b_2} + |H\rangle_{a_1}
|H\rangle_{b_1} |V\rangle_{a_2} |V\rangle_{b_2} + |V\rangle_{a_1}
|V\rangle_{b_1} |H\rangle_{a_2} |H\rangle_{b_2})|\alpha \rangle_a
|\alpha \rangle_b $ to the state
\begin{eqnarray}
&\rightarrow & (|H\rangle_{a_1} |H\rangle_{b_1} |H\rangle_{a_2}
|H\rangle_{b_2}  \nonumber\\
&& + |V\rangle_{a_1} |V\rangle_{b_1} |V\rangle_{a_2}
|V\rangle_{b_2}) |\alpha \rangle_a |\alpha \rangle_b \nonumber\\
&& + |H\rangle_{a_1} |H\rangle_{b_1} |V\rangle_{a_2}
|V\rangle_{b_2}|\alpha e^{i\theta} \rangle_a |\alpha e^{i\theta}
\rangle_b  \nonumber\\
&& + |V\rangle_{a_1} |V\rangle_{b_1} |H\rangle_{a_2}
|H\rangle_{b_2}|\alpha e^{-i\theta} \rangle_a |\alpha e^{-i\theta}
\rangle_b.
\end{eqnarray}
This measurement can not be accomplished in a deterministic way,
just in a nearly deterministic way. That is, Alice and Bob can not
obtain the state $|H\rangle_{a_1} |H\rangle_{b_1} |V\rangle_{a_2}
|V\rangle_{b_2} + |V\rangle_{a_1} |V\rangle_{b_1} |H\rangle_{a_2}
|H\rangle_{b_2}$ perfectly, which is different from that with
QND$_2$ in Sec. \ref{epis}.

In summary, we propose a different purification scheme based on
two PDC sources with cross-Kerr nonlinearities. The task of
entanglement purification can be completed with two steps in this
scheme. First, we provides a primary entanglement purification
protocol for PDC sources with QND detectors by transferring the
spatial entanglement of photon pairs to their polarization. In
this protocol, the QND detectors act as not only the role of CNOT
gates but also that of photon number detectors, which provides a
good way for the next process to purify the entanglement of the
photon pairs more as they make the photon pairs  equivalent to
those coming from two ideal sources. Compared with the Simon-Pan
protocol for PDC sources \cite{polarization}, this protocol does
not require sophisticated single-photon detectors and can
distinguish the number of the photons coming from the four modes.
This advantage makes the two parties have the ability to complete
the entanglement purification perfectly. In the second process for
entanglement purification, new QND detectors are designed to act
as the role of CNOT gates. This protocol does not require  CNOT
gates based on linear optical elements, but possesses the same
yield of photon pairs purified as the protocols \cite{C. H.
Benneet,D. Deutsch} with CNOT gates, double that of the PBS
protocol \cite{panjw}. As a perfect CNOT gate is far beyond what
is experimentally feasible with linear optical elements, this
protocol may be an optimal one.

\section*{ACKNOWLEDGEMENTS}

This work is supported by the National Natural Science Foundation
of China under Grant No. 10604008, A Foundation for the Author of
National Excellent Doctoral Dissertation of China under Grant No.
200723, and by Beijing Education Committee under Grant No.
XK100270454.

\end{document}